# Global technology access in biolabs - from DIY trend to an open source transformation

Tobias Wenzel[1*]

[1] Institute for Biological and Medical Engineering, Schools of Engineering, Medicine and Biological Sciences, Pontificia Universidad Católica de Chile

*Corresponding author

E-mail: tobias.wenzel@uc.cl

## Abstract

This article illustrates how open hardware solutions are implemented by researchers as a strategy to access technology for cutting-edge research. Specifically, it is discussed what kind of open technologies are most enabling in scientific environments characterized by economic and infrastructural constraints. It is demonstrated that do-it-yourself (DIY) technologies are already wide spread, in particular in countries with lower science funding, which in turn is the basis for the development of open technologies. Beyond financial accessibility, open hardware can be transformational to the technology access of laboratories through advantages in local production and direct knowledge transfer. Central drivers of the adoption of appropriate technologies in biolabs globally are open sharing, digital fabrication, local production, standard parts use, and detailed documentation.

## Technology access in biolabs

Most scientists who conduct experiments in research laboratories face resource constrains such as access to methods, technologies, equipment, and reagents. Given the vast complexity of biological systems and the demands on scientific rigor and speed, these technological constraints can quickly become a main challenge when doing research in an academic laboratory engaged in knowledge generation and publication in the life sciences - here referred to simply as "biolab". If well-funded research hubs struggle with technology access, how do research labs in scientific environments characterized by economic and infrastructural constraints, here referred to as 'low resource' settings deal with these challenges and still produce globally relevant research outputs? Do low-cost do-it-yourself (DIY) approaches and free and open source hardware (here 'open hardware' or 'open technology') improve researcher's technology access? The aim of this paper is to give an account of how DIY and open technologies are implemented and taken up by researchers globally, and what effect this has on researchers with restricted access to technologies. These low resource constraints, as we will see, extend beyond monetary cost considerations into a less rich research and technology ecosystem, and therefore solutions are not only driven by cost but also other factors such as local fabrication, access to parts and direct knowledge transfer. Much of the reasons driving technology adoption in low resource settings also apply to laboratories in general, where the unique advantages of open technology are used to address new custom research questions, apply existing methods in new settings, and to increase the accessibility of methods.

## DIY hardware provides access to technology

The idea of building your own equipment is as old as experimental sciences. However, in the context of current professionalised commercial research laboratory infrastructure, the concept of researchers building their own laboratory tools is being re-discovered. The Web of Science (search date 2$^{nd}$ of May 2022), for example, indexes 48,461 publications on 3D printing (73% in the last four years); 9,478 mentioning Arduino microcontrollers (55% last four years); 6,221 on Open Hardware (47% last four years); 5,320 mentioning the low-cost Raspberry Pi computer (66% last 4 years). All these are popular methods or tools for DIY instrumentation which see fast growth in publication number each year. Arduino co-founder Tom Igoe is an academic who teaches at the of School of Art at New York University. With the Arduino, he wanted a tool [1] that was cheaper and easier to use for his students to produce interactive digital technologies as part of their art, for example in 'making things talk' during theatre plays. The idea has led to sales of over 10 million Arduino Uno boards [2], a clear example of the spread of DIY and open source technology. Most researchers can benefit from easy-to-control tools and now Arduinos are even used for primary and elementary school teaching [3]. The portfolio of low-cost tools is mostly driven by the larger Maker movement outside of academia, where people are investing into their local production skills to create and customise technologies. An important aspect of the maker trend is the access to digital fabrication, which allows the interchange of increasingly complex digital designs that can be produced locally with an automated fabrication machine. Digital fabrication includes 3D printing, laser cutting, CNC-milling, plotter-cutting, the fabrication of electric circuit boards, and even microfluidic chips and DNA synthesis. Communal workshops such as FabLabs, MakerSpaces and others have sprung up around the globe to provide access to digital fabrication.

The impact of DIY technologies in research can also be assessed by analysing recent scientific publications. When it comes to scientific publications in general, a few countries dominate the global ranking of manuscripts authored, as illustrated in Figure 1 top left. These countries typically have a large density of scientists and industrial economies that produce complex goods, as can be seen in the right graph of Figure 1. But what about publications mentioning DIY tools? While 3D-printing or open hardware may be also referred to in an industrial or theoretical context, the Arduino and Raspberry Pi boards are specific tools more likely to be used in practice when part of the studies methodology. The later publications are here used as a proxy for studies using DIY hardware in general. When searching for the Arduino or Raspberry Pi keywords in the method section of articles, the publication distribution in Figure 1 bottom left shows us interesting differences in comparison to the overall publications. The countries visibly overrepresented among DIY studies, highlighted in black, tend to be located in regions of the world with less science resources. With little comparative data available on laboratory technology use, this data provides an indicator that DIY technology does indeed contribute to technology access in low resource countries. Unfortunately, many studies in biological sciences using such tools do not highlight this methodological approach in their articles, making contributions in biology less discoverable [4].

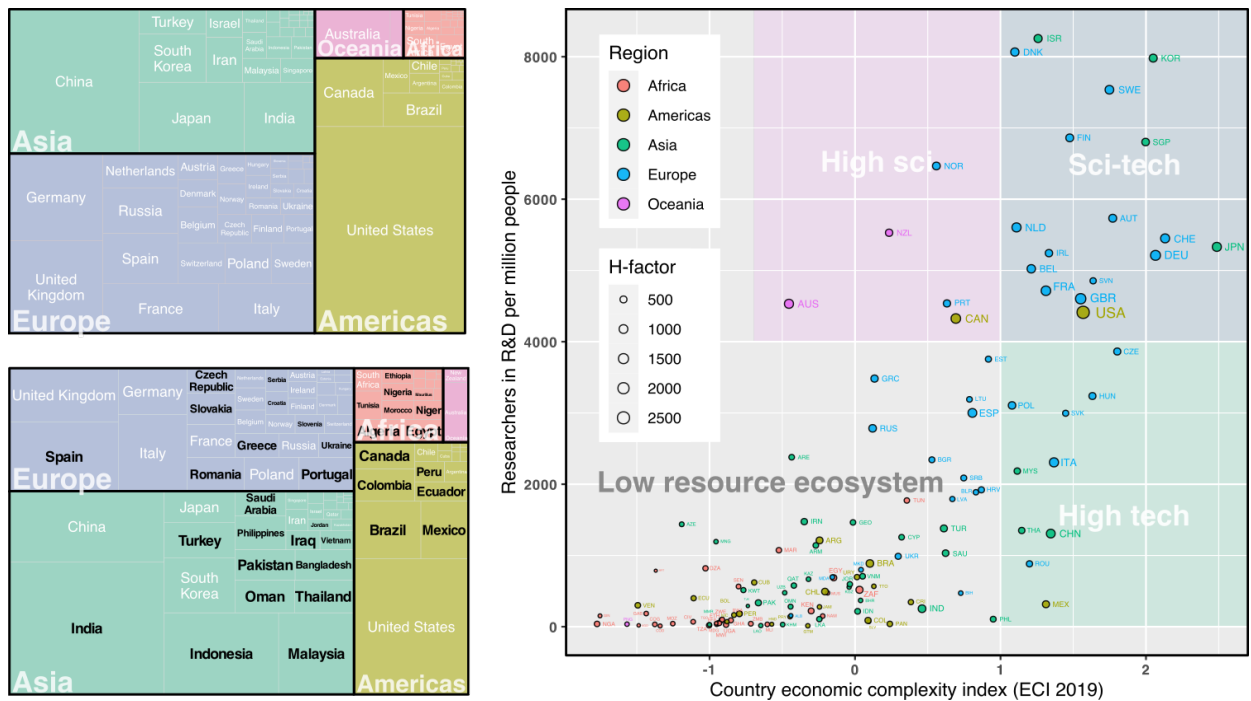

**Figure 1. Publication statistics for regional context of DIY technologies. (Top left)** Global distribution of total publications produced by country and regions. A country's fraction of total publications (58,485,740) is represented by its relative size in the graph. Data source: scimagojr total citable documents on any topic from 1996 to 2020. **(Bottom left)** Global distribution of publications that mention the tools 'Raspberry Pi' or 'Arduino' in the method section of the article by country and regions. A country's fraction of total publications (3718) is represented by its relative size in the graph. Countries visibly overrepresented in this category relative to overall publications are highlighted in black. Data source: Web of Science (WoS) search results on 'Raspberry Pi' or 'Arduino' across publications (articles, proceedings papers, early access, and data papers) indexed before the search date (15[th] of September 2022) – 13285 articles were found. This publication dataset was then compared to a Scite search on wider literature with 'Arduino' or 'Raspberry Pi' in the method section (23703 results). The datasets were then merged by DOI and where absent by title to obtain 3718 matches that contain the WoS publication type filter, method section criterion, and publication country data. The relative publication by country distribution for all WoS articles on the topic is similar to the result displayed for the method section only, see data repository. **(Right)** Graph showing each country (marked by country code and coloured by region) with respect to their density of researchers in the population (Y-axis) and their economic complexity index (ECI) (X-axis), while the size of the points represents the country's H-index. Data sources: researcher density - Worldbank Development Indicators, latest data available by country; ECI - Atlas of Economic Complexity, 2019 data or latest available; country H-index – scimagojr country rank, 1996-2020 all subjects; global regions – UN sustainable development goals; countries without data in all relevant datasets where removed.

This article reflects not only on tool access in low resource settings, but also on building and developing appropriate open tools by local laboratories. To this aim, international datasets are contrasted in the panel of Figure 1 Right, to contextualise low resource settings not only in terms of funding, but also in their access to scientific and developer know-how as well as industrial resources. The researcher density and economic complexity index [5] were chosen as quantifiable variables to represent a country's population-level 'high-science' or 'high-technology' environment respectively (or both or none). They illustrate the know-how and recourse context a scientist in the country might face when purchasing and recruiting to equip a laboratory.

## It's all about the money?

So, if laboratories in developing countries have less funding and are more likely to use low-cost DIY technologies, then is it simply a cost reduction strategy? The significant cost reductions that can be achieved with DIY and open source hardware are highlighted in the literature [6,7]. DIY hardware components often cost only 1%-10% of commercial proprietary tools, offering substantial savings for low-resource labs and enabling new applications where the equipment cannot be simply reused, for example, building a microscope to be placed inside an incubator with pathogens. This cost reduction makes some open technologies popular budget tools. However, there are several more key aspects that influence the choice of using open technologies, particularly knowledge transfer advantages and local fabrication.

Knowledge transfer in a research context is broadly defined as the dissemination of knowledge, ensuring that research insights are used in practical terms. However, knowledge transfer is often narrowly interpreted as commercialization through patents and companies holding licences. The recent growth of open hardware technologies has led to an alternative model [8], where open source designs allow a faster and earlier adoption of new methods by scientific users [9]. Open sourcing a design allows early adopters to build the new instrument before it might become commercially available, which in turn leads to timely feedback and improvements to the original design in some cases. The motivational aspect for developers to speed up adaptation of their technology is an important aspect to mention alongside simple cost reduction of hardware and is likely to contribute to the growing popularity of open technology among developers.

## Local fabrication and availability are key

Another key aspect of technology choice is local production. In any research lab, access to 3D printers and a good connection to a CNC-milling workshop can lead to faster instrumentation design turnaround time and a better ability to customize set-ups. For laboratories in low resource settings in particular, most suppliers and service providers are not easily available. Acquiring state of the art equipment in these places is much more expensive and takes much longer with uncertain shipping and import costs [10]. Furthermore, maintenance of such equipment, is often unsustainable due to costs and part-availability, missing specific know-how of local staff, and the expensive long flights between the laboratory and the company's maintenance team. These logistic and training challenges leave many instruments in low-resource countries out-of-order [11–13]. Therefore the DIY approach can be the only affordable and, if well documented, sustainable option for labs in scientific environments characterized by economic and infrastructural constraints. In this context, the local control of versatile digital means of fabrication can make an important difference to the technology that can be made available in a laboratory. For example, it allows researchers to 3D print the newest microscope stage model found online or emailed from overseas, using standard printing filament supplies and about two days of time, e.g. [14]. The global availability of the Arduino and Raspberry Pi boards, among others, is therefore also a likely contributor to their popularity for science in low resource settings. Not just the low cost of tools, but also their availability in the limited portfolio of local sales and their reusability in different experiments is relevant. DIY set-ups can also easily be automated and useful data export formats chosen without requiring specialized engineering knowledge. It must be noted that there are regions of the world where accessing a low-cost 3D printer and lab members to operate it may be a challenge. These extreme resource challenges may be partially mitigated through frugal open technology, training, and core facilities, but discussing it goes beyond the scope of this article.

The world has recently witnessed the importance of local production. With the beginning of the COVID-19 pandemic new medical needs suddenly emerged. In this moment of global shortage, many equipment producing countries first supplied their local needs, enforced by export banns, or not being able to deliver elsewhere through collapsed trade routes. In response, many people spontaneously developed DIY and open source solutions to locally produce personal protective equipment such as 3D printed face shields and even medical instrumentation, especially mechanical ventilators [15].

## The transformative potential of open hardware and wetware

In summary, it has been shown that DIY technology is widespread and used to access technology. Its advantages besides cost reduction are direct application, local fabrication and re-usability of parts when needed. But what exactly is the difference been DIY and 'open' technologies previously mentioned? Open source hardware is defined as a physical object whose design is made publicly available so that anyone can study, modify, distribute, make, and sell the design or hardware based on that design [16]. In contrast to the *practice* of do-it-yourself (DIY), open hardware is therefore a *concept* that describes how the design (inherently DIY in the eyes of the maker) is made available – at least with a documentation, modifiable design files and a suitable license. More about best practices (see OSHWA) and standardisation efforts [17] can be read elsewhere.

In practice, open technology can take a broad diversity of forms and shapes. Info Box 1 categorises open technology designs into 10 stereotypes to make sense of their differences in the context of this article. Even among researchers aiming to make their designs openly available, most end up not sharing design files for practical reasons (type 8, Info Box 1) [18] which makes these projects merely DIY, not open source. Open technologies span many 'maker style' life science projects as discussed so far (types 1-5 and 10) which are particularly 'make-able' when based on digital fabrication designs (types 1, sometimes 3, and 10). Open technologies even include biological wetware, which also consists of physical objects and can even be digitally fabricated (category 6), e.g. in the case of enzyme encoding genes. Open source can be low-cost and simple, but can also be very technically challenging and expensive, such as the CERN particle accelerator electronics (stereotype 9) or computer processors (stereotype 7). The later are usually unsuitable for biologist.

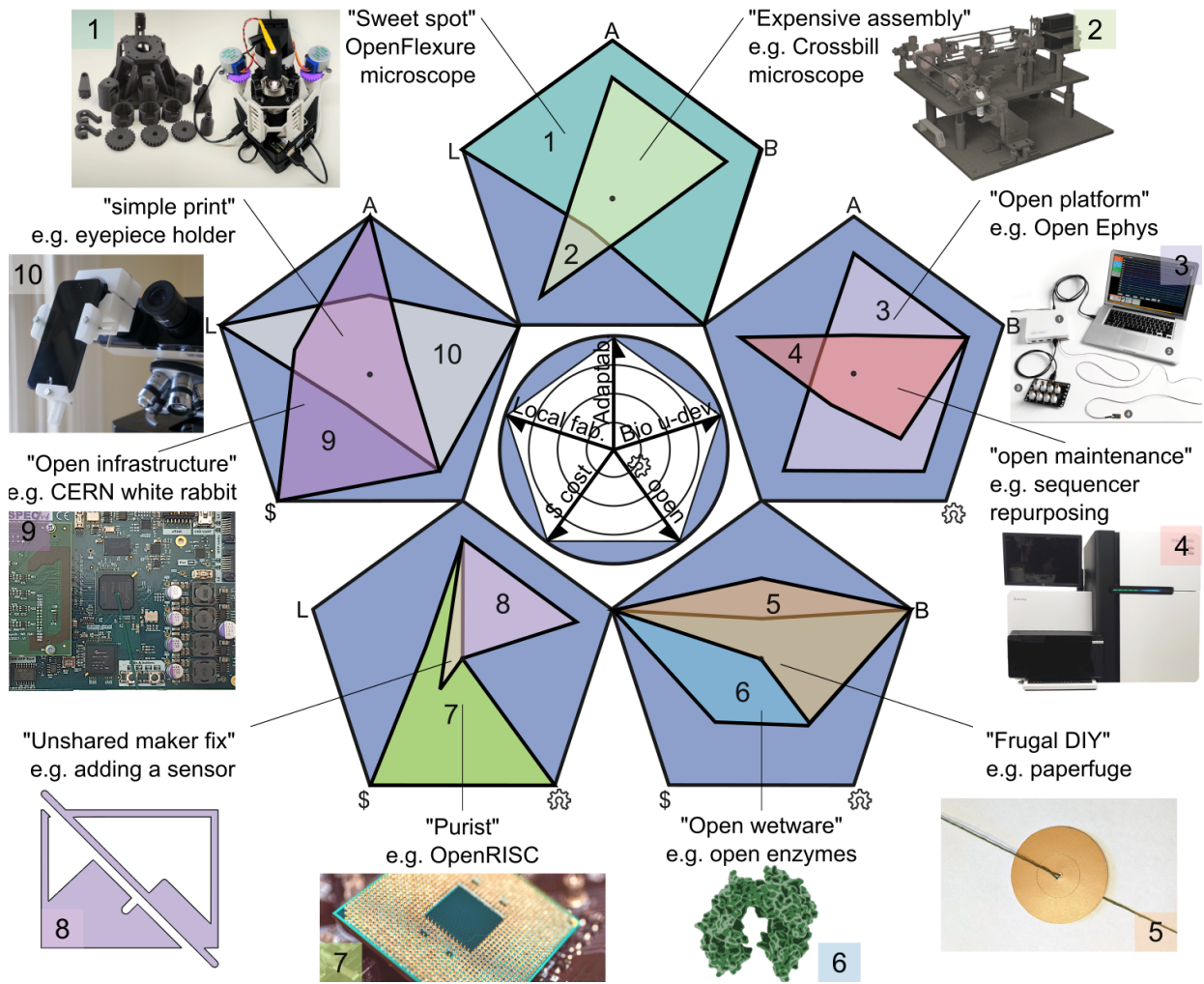

**Figure 2**. **Ten types of open hardware.** Illustration of the diversity of projects that fall within the portfolio of open hardware designs in a research context. Ten types are conceptualised, named, exemplified and rated along five dimensions (see scale in the centre): local fabrication suitability - can components be fabricated almost anywhere?; adaptability of the design; suitability for life science users to interact with the project documentation and contribute as developers; availability of modifiable open source design files; and the cost of needed components. The selected dimensions were rated subjectively to differentiate types in general, not only the given examples. See Box 1 for further explanation and references.

**Box 1**. **Different types of open hardware.** There are different types of open hardware projects for research, here categorised into ten groups shown in Figure 2. Each of them have their own

application advantages and challenges. In the (1) 'sweet spot', e.g. the OpenFlexure microscope [19], projects combine high qualities (including low-cost) of all five here-chosen categories as particularly important for biolab use and low-resource settings, but few such designs exist to date. Many laboratories use the ease and low-cost of the maker approach but never share solutions despite best intentions. (8) 'Unshared maker fixes' is therefore the most common category and also not actually open source. All other categories have the potential to meet the open source hardware definition. This open nature even applies to instructions that may not contain any modifiable design files but nevertheless open up hardware know-how, e.g. (2) 'expensive assemblies' which consist of proprietary parts sold by science equipment providers, such as optomechanical components from Thorlabs. These components are usually expensive and cannot be self-fabricated, but through a clever combination such equipment may give access to new methods and still save significant resources on a comparable ready-to-use equipment along with a reduced 'black-box character' [20]. Ready-to-use instruments may serve for modification or maintenance instructions in the (4) 'open maintenance' category, that can only be made with a specific proprietary instrument out of service, e.g. [21]. Such projects may not require additional hardware components, and when they do such components are often suitable for local fabrication such as 3D printing. (3) Science 'open platforms' such as Open Ephys [22] or OpenTrons [23] are modular and usually moderately priced and commercially available instruments that also avoid the science-unfriendly black-box character of other instruments. On the downside, their instruments are usually optimised for industrial production and therefore do not lend themselves to local fabrication or parts purchases. (5) 'Frugal DIY' projects claim the record for ultra-low cost [24], and can be made from locally available materials, at the same time they are usually limited in their applications and the lack of digital fabrication design limits their adaptability in research laboratories. They may also not comply with laboratory standard operating procedures, while on the other hand opening the door to science education for the general population. (6) 'Open enzymes' and other bioreagents, e.g. [25], are made available through an emerging effort to support the local production of such goods but doing so is currently challenging and batch quality an issue for further local distribution. All aforementioned stereotype projects are usually well suited for the participation of biologists in their development, along with one more category (10), the 'simple print', which typically does not require much documentation or fabrication effort, such as a phone holder, though this example is already comprised of several parts [26]. In contrast, (7) 'Purist' projects to open up elementary hardware such as computer processors are usually not found in life science projects. Last but not least, open hardware does not have to be low-cost and can be (9) 'open infrastructure' such as much of the electronic systems at CERN, e.g. [27,28]. This infrastructure has the advantage that its intellectual development and maintenance is driven by the organisation/public and does not break when proprietary providers change, but requires the collaboration with open developers and support companies. In the biolab context, infrastructure may manifest as equipment of type (2) 'expensive assemblies' in core facilities.

The open hardware concept is key to this discussion because it describes the difference between those internal designs in labs that simply take advantage of the ease of creating solutions 'maker style', and those that are shared in a useful way for others. Only open projects can be adapted to local needs and further developed to technology designs that lie beyond the development means of a single laboratory. The impressive growth potential of open source projects has many examples in the older movement of free and open source software, such as Linux and FIJI. Early examples of open technologies such as the OpenFlexure microscope (type 1 'sweet spot' in info box 1) provide us with a glimpse of where open hardware projects can go if they are built on digital design files and support the formation of a community. The OpenFlexure designs are open, low-cost, high-performance and can be automated e.g. for slide scanning. They are easy to adapt, and their designs use standard parts and modular digital files for local fabrication by 3D printing.

The mindset to use and build on each other's knowledge to advance science, or 'stand on the shoulder of giants' in Newton's words, is increasingly being applied again to biology laboratory instrumentation. Although the documentation of individual projects may not always be perfect and community standards are still evolving [17], it is recommended for laboratories to engage in this ongoing transformation, fail, learn and try again with more success as a means of benefitting from the advantages for experimental research. Resources are available on all levels [4].

## Local development is about appropriate technology not cheap hacks

It is necessary to discuss local fabrication and development not only in the context of its practicality, but also its meaning. Especially in the bio-sciences, new students might not be aware that building one's own tools has multiple advantages and is becoming a common practice in laboratories worldwide. As aforementioned, this does not mean that the instrument has a low-performance or is unreliable; rather, it is a useful strategy for technology control and innovation. Most labs that self-identify as technology development teams are based in countries with many scientists and a complex industry (see country codes in the top-right corner of the graph in Figure1 right), and have access to relatively large funding sources and industry collaborations. There, development projects are often undertaken with modular tools that are specifically commercialised for science developers and therefore look somehow 'professional'. This includes black-anodised optomechanical parts e.g. from Thorlabs, laser-modules with top-of-the-line branding, National Instruments automation boards running their proprietary modular software libraries, the latest use-case specific objective lens and neat commercial sample holders. And why not? It can be an appropriate approach when funding allows it, when there is stock at the local supplier and support teams offer customisation. More importantly, these tools tend to be reliable and compatible, and can thus save the team the time and hassle of making cheaper tools compatible as well as double checking their performance. Replicating instruments from such assembly guides (category 2 of info-box – "expensive assembly") may or may not be cheaper than turn-key commercial alternatives, and they may or may not contain other open source elements such as software or 3D design files. Still, useful instrumentation knowledge is disseminated and thanks to the modularity of tools, it can be further modified and improved.

Scientists, on the other hand, who develop open technologies with a more ambitious definition of accessibility, with globally available tools and at lower cost, continue to face important symbolic and institutional restrictions, despite the multiple benefits for scientific and technological development of such an approach, e.g. [29]. Such self-build are sometimes viewed as unprofessional or unsuccessful in grant applications and until recently faced barriers from academic journals to publish their work. Fortunately, currently there are many good and well-published examples of open hardware that can be used to underline the scientific vision of appropriate technology, e.g. [30–38]. Still this stigma has implications for sharing, as the developers might not feel that their solutions are proficient enough to share, and they might not self-identify as technology developers. The term 'appropriate technology' is useful to explain why many such solutions are nevertheless of interest to 'top-of-the-pyramid' scientists and 'lower resource labs' alike. "A technology is deemed to be appropriate when it is compatible with local, cultural, and economic conditions (i.e., the human, material and cultural resources of the economy), and utilizes locally available materials and energy resources, with tools and processes maintained and operationally controlled by the local population" [39] , and when possible open source [40]. Aforementioned tools that may be appropriate in a few rich and research focussed countries may be difficult to obtain, maintain and afford in most other countries, making technologies based on these tools inappropriate for research use elsewhere. Technology development labs that have access to a wide array of tools may use lower cost technology to enable new applications, or consider global access to components in order to improve opportunities for global adoption of their appropriate technology and thereby increase the impact of their methods. Either way, such technologies may simply be 'appropriate' for the task. For reasons of cost, access and operation it may not be appropriate to e.g. use professional chemically resistant microfilters to fish microplastics from the ocean off the shores of Newfoundland, and instead baby's tights may be the right tool for the job to do serios scientific research [41]. Digital fabrication is also an appropriate approach to build technology for research in a wide range of locations globally. It is therefore recommended to assess whether a tool is an appropriate technology, given its intended application context, rather than just highlighting its low cost or even using the negatively connotated term 'cheap'. If highlighted as appropriate technology, designs in manuscripts are less likely to miss their audience.

## Open technology trends in biolabs

Concretely, which are good current examples of appropriate open technology for biolabs? In recent years, a broad portfolio of open technology designs has been made available, enough to equip entire life science laboratories, as illustrated in Figure 3. These instruments are a good way to start engaging in open technologies, and there are many more helpful examples published, e.g. a 3d printable rotator mixer for incubators [42], a portable CO2 incubator for tissue cultures [43], a 3D printable spectrophotometer [44], an anaerobic chamber with DIY catalyst [45], a system for automated parallel microbial cultivation [35], an isothermal well-plate reader for LAMP reactions 'MIRIAM', an open source Prusa 3D printer modified for bioprinting [46], and many more. Besides integrated and automated equipment, there is a wealth of simpler open source 3D printable designs such as equipment adapters, covers, clips, sample or pipette holders, well-plate locator stands, physiological models and more, see Thingiverse, Instructables, Hackster.io, Wikifactory and other platforms. These can be re-designed or adapted to local needs. For example this semester some of my students (course IIBM2026 at UC Chile) designed 3D printed pipette holders that hold pipettes in place with a gentle flex-mechanism (see also Figure 3), so that the pipettes are less likely to fall during Chile's frequent shakes.

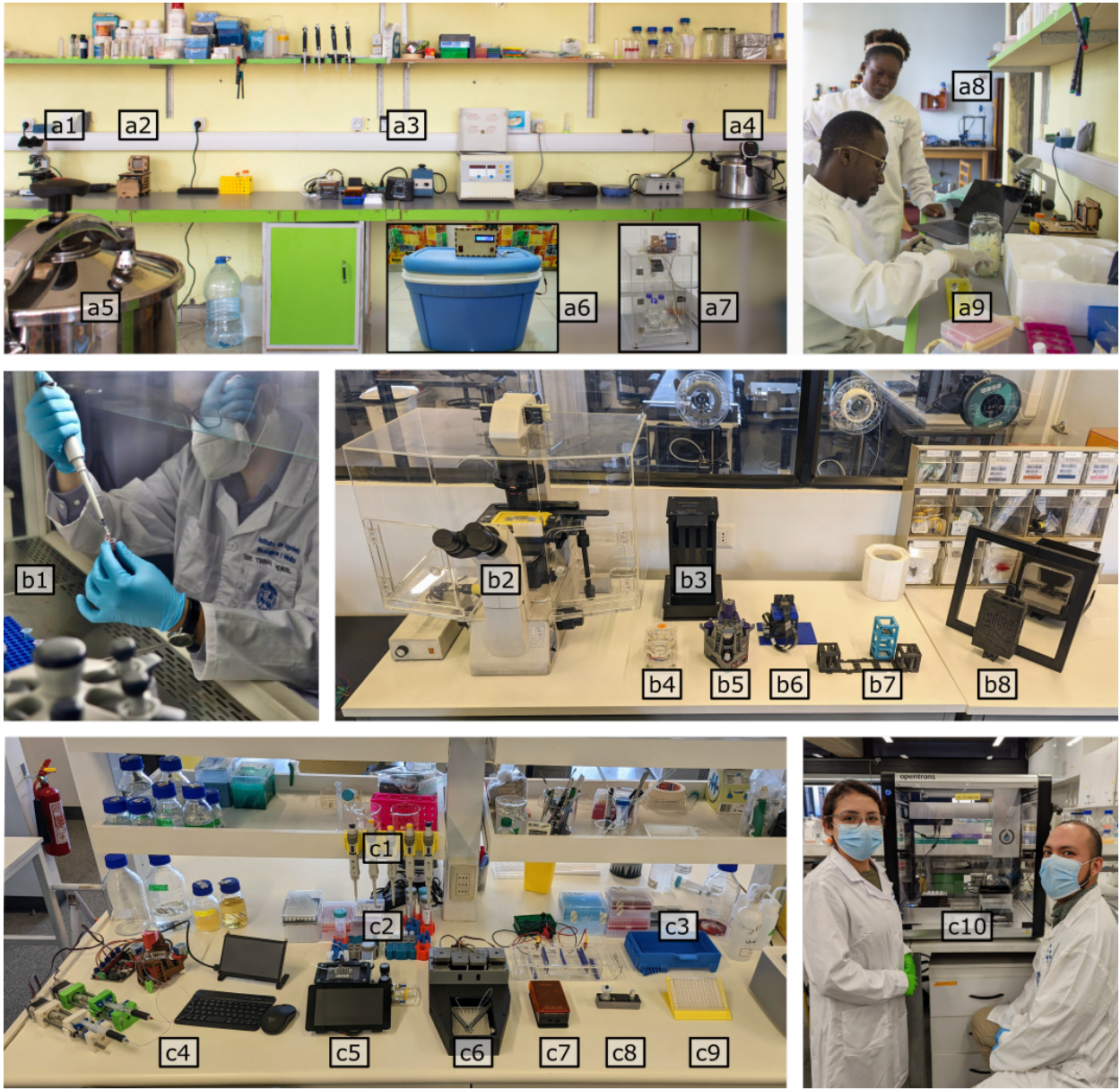

**Figure 3: Southern bio-laboratories equipped with open technology.** The top images show the Mboa Lab in Cameroon (source: openbioeconomy.org) and middle and lower images show the Wenzel Lab in Chile. The following equipment is shown: (a1) basic commercial light microscope, (a2) open hardware thermocycler – rebuild of OpenPCR, (a3) small commercial DNA analysis equipment - miniPCR, bluegel, vortexter, centrifuge, (a4) repurposed Sous Vide as water bath, (a5) repurposed pressure cooker as autoclave, (a6) self-designed open hardware incubator, (a7) self-designed open hardware shaking incubator, (a8) commercial open hardware 3D printer, (a9) use of reagents and open enzymes. (b1) Second hand commercial laminar flow hood, (b2)

second hand commercial microscope with custom enclosure and 3D printed sample and camera adaptors, (b3) open source laser-cut fluorescence plate imaging station – rebuilt [51], (b4) open laser-cut smartphone microscope - Roachscope from Backyard Brains, (b5) open 3D printed OpenFlexure microscope in fluorescence configuration – rebuilt and star-LED fitted [50], (b6) 3D printed computational microscope – rebuilt with Unicorn Raspberry Pi hat, (b7) open 3D printed holographic microscope - UC2 rebuild [30], (b8) open, time-laps, multiple Petri plate imaging system – SPIRO rebuild in progress [52]. (c1) Student-designed 3D printed pipette holders with earthquake safety, (c2) open 3D printed tube racks – rebuilt from Dormant Biology Lab design, (c3) 3D printed pipette tip aligner for tip box refill – Elster rebuild, (c4) self-designed microfluidics educational workstation – Raspberry Pi based control of 3D printed syringe pumps and microscope, (c5) open workstation for integrated single cell transcriptomics – RNA-seq miniDrops rebuild [32], (c6) open 3D printed microscopy pipetting robot that can be mounted on a microscope – rebuild in progress [31], (c7) open source gel electrophoresis set - self-designed parametric laser-cut gel chamber with open commercial IORodeo power supply and IORodeo blue LED transilluminator, (c8) open thermocycler - Gaudilab PocketPCR built from commercial set, (c9) open self-designed well-plate locator stand, (c10) open commercial pipetting robot OpenTrons.

Here, general purpose equipment is highlighted for broad interest but many of these instruments also have several commercial & proprietary alternatives. At the same time, open solutions particularly stand out at solving special applications that few other people currently share, or where equipment cannot be recovered, e.g. outdoor sensors and tech for infection research such as COVID samples, etc. [50,53]. Another appropriate approach for biological experiments is to use low-cost electronic boards to automate or add sensors to experiments, e.g. for humidity, pressure, temperature or positioning. In this context, it often does not matter to biologists if the board itself is open source (e.g. Arduino) or not (e.g. Raspberry Pi) – it is simply used as a component in an otherwise hopefully well-documented set-up. Biology examples and approaches for open electronics have recently been described elsewhere [4]. Two particular active areas of open development are microscopy [54] and neuroscience (see commercial portfolios in Table 1). Another exciting area of growth is open wetware which democratises access to enzymes, reagents, and reagent fabrication designs [25,55,56]. Automation for sample handling and calibration is furthermore used to improve reproducibility of research.

Currently, it is rare to see a laboratory with many open technologies used at the same time. Usually, even in low-resource settings, laboratories are a mix of some relatively new instruments funded by research grants, second hand equipment (usually older, borrowed from colleagues, moved from other laboratories or projects, and some donated), and self-made solutions starting from buffer-mixes, coat hangers and incubator racks, to more complex research instrumentation as discussed above. The mix of technology types in use is also illustrated in the laboratory examples from Africa and Latin America in Figure 3. Expensive set-ups can also be an option in low resource settings, e.g. in the context of core facilities when made available to other research users that pay for use or local service.

## From the lab to industry

Open technology is not the opposite of commercialisation. Engaging in open technology development does not mean to oppose the mainstream approach of knowledge transfer through commercialisation. As in the example multi-billion dollar open software companies, non-proprietary products can still be sold, marketed, developed and serviced. All these are valuable ways of making an open design even more available. Open hardware related business models are an ongoing area of development [57,58] and practise, demonstrated by the list of open hardware company examples in Table 1.

**Table 1: Bio-laboratory related open hardware company examples**

| Company name | Field or product |
|---:|---|
| Tindie | online marketplace for electronics and hardware products. Sales platform for makers and small businesses |
| SparkFun | designs, manufactures and sells microcontroller development boards and breakout boards |
| Seedstudio | integrated platform for creating hardware solutions for IoT, AI, and Edge |
| Sci-bots | Dropbot, digital microfluidic control system |
| Sanworks | automated systems for Neuroscience research |
| Safecast | crowdsourced environmental data with open hardware, originally radiation in Japan now also air quality globally |
| RedPitaya | FPGA-powered boards as multifunction laboratory and engineering instrument |
| Public Lab | community science tools for environmental monitoring |
| Prometheus Sci | open hardware for biology consulting |
| Precious Plastic | plastic recycling equipment |
| OpenTrons | pipetting robot and addons |
| OpenQCM | Quartz Crystal Microbalance, e.g. for protein affinity measurements at molecular resolution |
| OpenFlexure Industries | manufactures the OpenFlexure microscope and OpenFlexure block stage |
| OpenEphys | electrophysiology laboratory tools |
| OpenBCI | brain-computer interface platform |
| Open source commercial 3D printers | Prusa printers (Prusa Research), Lulzbot (Frame 3D), Voron, Rostock Max (SeeMeCNC), BigFDM |
| Open Acoustic Devices | designs supports and deploys acoustic hardware and software for environmental and wildlife monitoring projects, audio moth |
| Olimex | developer and provider for development tools for embedded market, including non-open boards |
| neurogig | instruments for neurophysiology laboratories |
| Kitspace | share and order electronic projects with the automatic component shopping list generator for GitHub projects |
| IORodeo | designing and selling bio-lab hardware, now mainly potentiostats |
| Elphel | Imaging hardware and software |
| Beneficial Bio | network of social enterprises for biolabs |
| Beagle Board | BeagleBone computers |
| BackyardBrains | demonstration instruments for neuroscience education |
| Arduino | designs and manufactures single-board microcontrollers and microcontroller kits for building digital devices |
| Apertus | AXIOM cameras |
| Adafruit | designs, manufactures and sells a number of electronics products, electronics components, tools and accessories |

Often academics and universities do not have the means, legal or industrial knowledge to enforce global intellectual property infringements. Publishing them openly on the other hand can protect from future patent claims by others. And while patents are still a good marketing tool to investors, the several year cycle of patent processing is often out of sync with fast-moving technological development. The investing markets might also not be where local problems could be solved with different business models and without the pressure of venture capital. Open projects can be faster, more flexible, and perhaps most importantly, closer to the users. Users can directly modify and improve the open technology at hand, thus both advance technologies, contribute to local availability of solutions, and empower citizens [59–61].

## Open transformation barriers

As discussed, biolaboratories in low resource settings face challenges around technology access that are far more complex than the availability of money to buy instruments. Open source designs and digital local production are increasingly employed as promising solutions to several of these challenges. So why is open technology not yet actively supported by funders and institutions in these regions? There is no conclusive data available to answer this question rigorously. The following points might contribute to understand the status quo: (i) open hardware sharing is a relatively recent concept and not yet known by all; (ii) scientific development in developing countries tends to follow the patterns of scientific development in developed countries where standards of open hardware have rarely been implemented at an institutional level, but first examples are emerging [62] (iii) lower resource countries tend to have a lower density of scientists and a less complex economy and therefore less complex local production facilities. This is demonstrated graphically in Figure 1 (right), where most lower-resource countries cluster in the bottom-left section of the graph. This lacking technology development ecosystem results into less access to production resources, and less access to qualified personal to build instruments. In fact, fabrication skills are highly sought after and knowledgeable individuals can contribute to the migration of advanced human capital away from low-resource settings. This insight into the local ecosystem highlights the need for detailed documentation and the use of standardised and accessible tools to enhance regional application. (iv) Many lower-resource countries and laboratories do not see themselves as technology developers, and therefore perhaps do not consider their local solutions as worth sharing. On the contrary, in many places foreign and western brands are highly valued, which can extend to laboratories and their PIs wanting to be seen as professionals working with popular international equipment. (v) In China [63,64] and some parts of Asia complex production facilities and technical know-how are available and enable fast and low-cost prototyping of technology, but designs and calibration instructions are not often shared in detail. The lack of documentation sharing may be caused by a believe that people anywhere could easily build something equivalent, or a different mindset around intellectual property licensing, technology transfer, task-sharing, communication and documentation practices.

We could see an even faster growth of custom life science technology designs if we adjusted institutional policies [62] to incentivise departmental workshops to release the frequent custom designs [18] they create for individual researchers. Often the opposite is the case and institutional workshop staff is not allowed to release files or hand the design to the respective academic to make the files available. Sometimes academics are also prevented from open sourcing their potentially patentable developments. This is one more reason to work openly from the start of a project and release documentation as the research develops. This allows early collaboration, feedback, and with this approach there is no point in time at which an institutionally problematic amount of intellectual property is released. The 'open from the start' model also avoids the back lock of documentation tasks left for final deadlines.

## Conclusion

Open technology can be a powerful strategy to access appropriate research technology in both low- and high-resource biolabs. This fast-growing strategy generally does not aim to break with the tradition of proprietary and commercial approaches, instead it adds a new dimension to current mainstream knowledge transfer models. Supporting openness can lead to a transformative increase in technology access and local development, especially in settings which are currently challenged in this regard. The new, fast, user-centred, open approach promises a place at the 'global research table' with a more widespread access to appropriate research technologies; locally developing, adapting, and using technologies appropriate for local contexts and enhancing regional innovation and business. A large portfolio of designs already exists with a broad spectrum of design approaches, cost, commercial availability and complexity – with something to offer for anyone who would like to become an open technology user and developer. If we limit ourselves to the current mainstream practise of low-resource laboratories to mainly buy or inherit old, basic or hardly maintainable equipment, regional bio-researchers will most likely keep living in the scientific past and miss the opportunity of being at the forefront of scientific development in the future. Every user and builder is a local gain of good examples, technological know-know and infrastructure enablers, in ways that proprietary closed technology does not enrich the local environment after purchase and shipping, since their technology is not open to be locally known, built and evolved. Future changes in relevant institutional guidelines and funders policies worldwide should aim to recognise, encourage and enable the open technology approach. Such institutional recognition and the availability even of relatively small flexible funds can make a large difference to laboratories assembling open technology.

## Acknowledgements

The author thanks Martina Yopo Diaz for helpful comments on the manuscript, Pierre Padilla Huamantinco for help preparing for the laboratory photo, and many of his colleagues and the open hardware community for discussions. He is grateful for funding from ANID FONDECYT Iniciacion 11200666 and the CZI project 'Latin American Hub for Bioimaging Through Open Hardware'. The funders had no role in study design, data collection and analysis, decision to publish, or preparation of the manuscript.

## Data Availability

The R code to generate the panels of Figure 1 is shared in the following repository: [https://osf.io/59bv2/](https://osf.io/59bv2/). Data sourced from public repositories is also shared there along with processed datasets.

# References

1. Igoe T. Interview Session: Tom Igoe, co-founder of Arduino; https://perma.cc/2BLT-A5XR. 2015. Available: https://medium.com/@_peterryan/interview-session-tom-igoe-co-founder-of-arduino-717c71ed6608

2. News. 10M Arduino Uno boards sold worldwide. Control Design; https://perma.cc/PX4F-M58X. 9 Dec 2021. Available: https://www.controldesign.com/industrynews/2021/10m-arduino-uno-boards-sold-worldwide/

3. Lee E. A Meta-Analysis of the Effects of Arduino-Based Education in Korean Primary and Secondary Schools in Engineering Education. European J Educ Res. 2020;9: 1503–1512. doi:10.12973/eu-jer.9.4.1503

4. Oellermann M, Jolles JW, Ortiz D, Seabra R, Wenzel T, Wilson H, et al. Open Hardware in Science: The Benefits of Open Electronics. Integr Comp Biol. 2022. doi:10.1093/icb/icac043

5. Hidalgo CA. Economic complexity theory and applications. Nat Rev Phys. 2021;3: 92–113. doi:10.1038/s42254-020-00275-1

6. Pearce JM. Economic savings for scientific free and open source technology: A review. Hardwarex. 2020;8: e00139. doi:10.1016/j.ohx.2020.e00139

7. Chagas AM, Prieto-Godino LL, Arrenberg AB, Baden T. The €100 lab: A 3D-printable open-source platform for fluorescence microscopy, optogenetics, and accurate temperature control during behaviour of zebrafish, Drosophila, and Caenorhabditis elegans. Plos Biol. 2017;15: e2002702. doi:10.1371/journal.pbio.2002702

8. Fantner GE, Oates AC. Instruments of change for academic tool development. Nat Phys. 2021;17: 421–424. doi:10.1038/s41567-021-01221-3

9. Power RM, Huisken J. Putting advanced microscopy in the hands of biologists. Nat Methods. 2019;16: 1069–1073. doi:10.1038/s41592-019-0618-1

10. Helden P van. The cost of research in developing countries. Embo Rep. 2012;13: 395–395. doi:10.1038/embor.2012.43

11. Malkin R, Keane A. Evidence-based approach to the maintenance of laboratory and medical equipment in resource-poor settings. Med Biol Eng Comput. 2010;48: 721–726. doi:10.1007/s11517-010-0630-1

12. Webber CM, Martínez-Gálvez G, Higuita ML, Ben-Abraham EI, Berry BM, Porras MAG, et al. Developing Strategies for Sustainable Medical Equipment Maintenance in Under-Resourced Settings. Ann Glob Health. 2020;86: 39. doi:10.5334/aogh.2584

13. Perry L, Malkin R. Effectiveness of medical equipment donations to improve health systems: how much medical equipment is broken in the developing world? Med Biol Eng Comput. 2011;49: 719–722. doi:10.1007/s11517-011-0786-3

14. Stirling J, Sanga VL, Nyakyi PT, Mwakajinga GA, Collins JT, Bumke K, et al. The OpenFlexure Project. The technical challenges of Co-Developing a microscope in the UK and Tanzania. 2020 Ieee Global Humanit Technology Conf Ghtc. 2020;00: 1–4. doi:10.1109/ghtc46280.2020.9342860

15. Chagas AM, Molloy JC, Prieto-Godino LL, Baden T. Leveraging open hardware to alleviate the burden of COVID-19 on global health systems. Plos Biol. 2020;18: e3000730. doi:10.1371/journal.pbio.3000730

16. OSHWA OSHA. Open Source Hardware Definition. 11 Feb 2011 [cited 27 Jun 2022]. Available: https://www.oshwa.org/definition/

17. Bonvoisin J, Molloy J, Häuer M, Wenzel T. Standardisation of Practices in Open Source Hardware. Journal of Open Hardware. 2020. doi:10.5334/joh.22

18. Diederich B, Müllenbroich C, Vladimirov N, Bowman R, Stirling J, Reynaud EG, et al. CAD we share? Publishing reproducible microscope hardware. Nat Methods. 2022; 1–5. doi:10.1038/s41592-022-01484-5

19. Sharkey JP, Foo DCW, Kabla A, Baumberg JJ, Bowman RW. A one-piece 3D printed flexure translation stage for open-source microscopy. Rev Sci Instrum. 2016;87: 025104. doi:10.1063/1.4941068

20. Kumar M, Kishore S, McLean DL, Kozorovitskiy Y. Crossbill: an open access single objective light-sheet microscopy platform. Biorxiv. 2021; 2021.04.30.442190. doi:10.1101/2021.04.30.442190

21. Pandit K, Petrescu J, Cuevas M, Stephenson W, Smibert P, Phatnani H, et al. An open source toolkit for repurposing Illumina sequencing systems as versatile fluidics and imaging platforms. Sci Rep-uk. 2022;12: 5081. doi:10.1038/s41598-022-08740-w

22. Siegle JH, López AC, Patel YA, Abramov K, Ohayon S, Voigts J. Open Ephys: an open-source, plugin-based platform for multichannel electrophysiology. J Neural Eng. 2017;14: 045003. doi:10.1088/1741-2552/aa5eea

23. Eggert S, Mieszczanek P, Meinert C, Hutmacher DW. OpenWorkstation: A modular open-source technology for automated in vitro workflows. Hardwarex. 2020;8: e00152. doi:10.1016/j.ohx.2020.e00152

24. Bhamla MS, Benson B, Chai C, Katsikis G, Johri A, Prakash M. Hand-powered ultralow-cost paper centrifuge. Nat Biomed Eng. 2017;1: 0009. doi:10.1038/s41551-016-0009

25. Matute T, Nuñez I, Rivera M, Reyes J, Blázquez-Sánchez P, Arce A, et al. Homebrew reagents for low-cost RT-LAMP. J Biomol Techniques Jbt. 2021;32: 114–120. doi:10.7171/jbt.21-3203-006

26. Vera RH, Schwan E, Fatsis-Kavalopoulos N, Kreuger J. A Modular and Affordable Time-Lapse Imaging and Incubation System Based on 3D-Printed Parts, a Smartphone, and Off-The-Shelf Electronics. Plos One. 2016;11: e0167583. doi:10.1371/journal.pone.0167583

27. Lipiński M, Bij E van der, Serrano J, Włostowski T, Daniluk G, Wujek A, et al. White Rabbit Applications and Enhancements. 2018 Ieee Int Symposium Precis Clock Synchronization Meas Control Commun Ispcs. 2018;00: 1–7. doi:10.1109/ispcs.2018.8543072

28. Wallbank JV, Amodeo M, Beaumont A, Buzio M, Capua VD, Grech C, et al. Development of a Real-Time Magnetic Field Measurement System for Synchrotron Control. Electronics. 2021;10: 2140. doi:10.3390/electronics10172140

29. Bezuidenhout L, Stirling J, Sanga VL, Nyakyi PT, Mwakajinga GA, Bowman R. Combining development, capacity building and responsible innovation in GCRF-funded medical technology research. Dev World Bioeth. 2022. doi:10.1111/dewb.12340

30. Diederich B, Lachmann R, Carlstedt S, Marsikova B, Wang H, Uwurukundo X, et al. A versatile and customizable low-cost 3D-printed open standard for microscopic imaging. Nat Commun. 2020;11: 5979. doi:10.1038/s41467-020-19447-9

31. Dettinger P, Kull T, Arekatla G, Ahmed N, Zhang Y, Schneiter F, et al. Open-source personal pipetting robots with live-cell incubation and microscopy compatibility. Nat Commun. 2022;13: 2999. doi:10.1038/s41467-022-30643-7

32. Stephenson W, Donlin LT, Butler A, Rozo C, Bracken B, Rashidfarrokhi A, et al. Single-cell RNA-seq of rheumatoid arthritis synovial tissue using low-cost microfluidic instrumentation. Nat Commun. 2018;9: 791. doi:10.1038/s41467-017-02659-x

33. Ambrose B, Baxter JM, Cully J, Willmott M, Steele EM, Bateman BC, et al. The smfBox is an open-source platform for single-molecule FRET. Nat Commun. 2020;11: 5641. doi:10.1038/s41467-020-19468-4

34. Kong DS, Thorsen TA, Babb J, Wick ST, Gam JJ, Weiss R, et al. Open-source, community-driven microfluidics with Metafluidics. Nat Biotechnol. 2017;35: 523–529. doi:10.1038/nbt.3873

35. Wong BG, Mancuso CP, Kiriakov S, Bashor CJ, Khalil AS. Precise, automated control of conditions for high-throughput growth of yeast and bacteria with eVOLVER. Nat Biotechnol. 2018;36: 614–623. doi:10.1038/nbt.4151

36. Pearce JM. Cut costs with open-source hardware. Nature. 2014;505: 618–618. doi:10.1038/505618d

37. Salazar-Serrano LJ, Torres JP, Valencia A. A 3D Printed Toolbox for Opto-Mechanical Components. Plos One. 2017;12: e0169832. doi:10.1371/journal.pone.0169832

38. Mulberry G, White KA, Vaidya M, Sugaya K, Kim BN. 3D printing and milling a real-time PCR device for infectious disease diagnostics. Plos One. 2017;12: e0179133. doi:10.1371/journal.pone.0179133

39. Hazeltine B. Field Guide to Appropriate Technology. Hazeltine B, Bull C, editors. Elsevier Inc.; 2003. doi:https://doi.org/10.1016/B978-0-12-335185-2.X5042-6

40. Pearce JM. The case for open source appropriate technology. Environ Dev Sustain. 2012;14: 425–431. doi:10.1007/s10668-012-9337-9

41. Liboiron M. Compromised Agency: The Case of BabyLegs. Engaging Sci Technology Soc. 2017;3: 499–527. doi:10.17351/ests2017.126

42. Dhankani KC, Pearce JM. Open source laboratory sample rotator mixer and shaker. Hardwarex. 2017;1: 1–12. doi:10.1016/j.ohx.2016.07.001

43. Arumugam A, Markham C, Aykar SS, Pol BVD, Dixon P, Wu M, et al. PrintrLab incubator: A portable and low-cost CO2 incubator based on an open-source 3D printer architecture. Plos One. 2021;16: e0251812. doi:10.1371/journal.pone.0251812

44. Laganovska K, Zolotarjovs A, Vázquez M, Donnell KM, Liepins J, Ben-Yoav H, et al. Portable low-cost open-source wireless spectrophotometer for fast and reliable measurements. Hardwarex. 2020;7: e00108. doi:10.1016/j.ohx.2020.e00108

45. Herrmann AJ, Gehringer MM. A low-cost automized anaerobic chamber for long-term growth experiments and sample handling. Hardwarex. 2021;10: e00237. doi:10.1016/j.ohx.2021.e00237

46. Krige A, Haluška J, Rova U, Christakopoulos P. Design and implementation of a low cost bio-printer modification, allowing for switching between plastic and gel extrusion. Hardwarex. 2021;9: e00186. doi:10.1016/j.ohx.2021.e00186

47. Poór VS. A low-cost and open-source mini benchtop centrifuge for molecular biology labs. Hardwarex. 2022;12: e00328. doi:10.1016/j.ohx.2022.e00328

48. Pitrone PG, Schindelin J, Stuyvenberg L, Preibisch S, Weber M, Eliceiri KW, et al. OpenSPIM: an open-access light-sheet microscopy platform. Nat Methods. 2013;10: 598–599. doi:10.1038/nmeth.2507

49. Chen B-C, Legant WR, Wang K, Shao L, Milkie DE, Davidson MW, et al. Lattice light-sheet microscopy: Imaging molecules to embryos at high spatiotemporal resolution. Science. 2014;346: 1257998. doi:10.1126/science.1257998

50. Collins JT, Knapper J, Stirling J, Mduda J, Mkindi C, Mayagaya V, et al. Robotic microscopy for everyone: the OpenFlexure microscope. Biomed Opt Express. 2020;11: 2447–2460. doi:10.1364/boe.385729

51. Nuñez I, Matute T, Herrera R, Keymer J, Marzullo T, Rudge T, et al. Low cost and open source multi-fluorescence imaging system for teaching and research in biology and bioengineering. Plos One. 2017;12: e0187163. doi:10.1371/journal.pone.0187163

52. Ohlsson JA, Leong JX, Elander PH, Dauphinee AN, Ballhaus F, Johansson J, et al. SPIRO – the automated Petri plate imaging platform designed by biologists, for biologists. Biorxiv. 2021; 2021.03.15.435343. doi:10.1101/2021.03.15.435343

53. Beddows PA, Mallon EK. Cave Pearl Data Logger: A Flexible Arduino-Based Logging Platform for Long-Term Monitoring in Harsh Environments. Sensors Basel Switz. 2018;18: 530. doi:10.3390/s18020530

54. Hohlbein J, Diederich B, Marsikova B, Reynaud EG, Holden S, Jahr W, et al. Open microscopy in the life sciences: Quo Vadis? Arxiv. 2021.

55. Gomez-Marquez J, Hamad-Schifferli K. Distributed Biological Foundries for Global Health. Adv Healthc Mater. 2019;8: 1900184. doi:10.1002/adhm.201900184

56. Kahl L, Molloy J, Patron N, Matthewman C, Haseloff J, Grewal D, et al. Opening options for material transfer. Nat Biotechnol. 2018;36: 923–927. doi:10.1038/nbt.4263

57. Pearce JM. Emerging Business Models for Open Source Hardware. Journal of Open Hardware. 2017. doi:10.5334/joh.4

58. Li Z, Seering W. Does Open Source Hardware Have a Sustainable Business Model? An Analysis of Value Creation and Capture Mechanisms in Open Source Hardware Companies. Proc Des Soc Int Conf Eng Des. 2019;1: 2239–2248. doi:10.1017/dsi.2019.230

59. Rey-Mazón P, Keysar H, Dosemagen S, D'Ignazio C, Blair D. Public Lab: Community-Based Approaches to Urban and Environmental Health and Justice. Sci Eng Ethics. 2018;24: 971–997. doi:10.1007/s11948-018-0059-8

60. Kera D. Science Artisans and Open Science Hardware. Bulletin Sci Technology Soc. 2017;37: 97–111. doi:10.1177/0270467618774978

61. Pearce JM. Distributed Manufacturing of Open Source Medical Hardware for Pandemics. J Manuf Mater Process. 2020;4: 49. doi:10.3390/jmmp4020049

62. EMBL. EMBL Policy - Open Science and Open Access. 2021 Dec. Available: https://www.embl.org/documents/document/internal-policy-no-71-open-science-and-open-access/

63. Huang AB. The Hardware Hacker: Adventures in Making and Breaking Hardware. Press NS, editor. No Starch Press; 2019.

64. Lindtner S. Hacking with Chinese Characteristics. Sci Technology Hum Values. 2015;40: 854–879. doi:10.1177/0162243915590861